\documentclass{osa-article}

\journal{oe}
\articletype{Research Article}
\begin{document}

\title{Microcavity resonance condition, quality factor, and mode volume are determined by different penetration depths}

\author{C. Koks,\authormark{1} M.P. van Exter,\authormark{1,*}}

\address{\authormark{1}Huygens-Kamerlingh Onnes Laboratory, Leiden University, P.O. Box 9504, 2300 RA Leiden, The Netherlands\\}

\email{\authormark{*}exter@physics.leidenuniv.nl} %% email address is required

\begin{abstract}
The penetration depth in a Distributed Bragg Reflector (DBR) co-determines the resonance condition, quality factor, and mode volume of DBR-based microcavities.
Recent studies have used an incomplete description of the penetration depth and incorrect equations.
We present a complete analysis that involves three different penetration depths. 
We also present a series of experiments on microcavities to accurately determine the frequency and modal penetration depth of our DBRs and compare these results with theoretical predictions.
The obtained results are relevant for anyone who models a DBR as an effective hard mirror if lengths of the order of the wavelength are relevant, as is the case for microcavities.
\end{abstract}

\section{Introduction}
\label{sec:introduction}
Small mode volume cavities have been used for numerous applications such as quantum information processing with individual atoms \cite{Bogdanovic2017} and lab-on-a-chip sensors \cite{Trichet2014}. 
These microcavities typically consist of two highly reflective Distributed Bragg Reflectors (DBR) which can trap light in a small mode volume and thereby increase the light-matter interaction. 
When microcavities get smaller \cite{Dolan2018}  the penetration depth in the mirrors becomes important.
DBRs are also used in many other applications \cite{Royo2002} and even exist in nature, in the form of intricate photonic crystals \cite{Barry2020}.

DBR-based microcavities are often modeled as cavities with two hard mirrors spaced by a cavity length that is extended by the penetration depths of the DBRs.
This model is then used to calculate the resonance condition, quality factor and mode volume.
However, the optical penetration in the DBRs is more subtle than this simple model suggests.
In the literature, the penetration depth in DBRs is ambiguously defined due to this simplified model \cite{Mader2015,Trichet2014, Greuter2014,Trichet2018,Vogl2019, Benedikter2019,  Benedikter2015}.

This paper will solve these issues by introducing multiple (frequency, modal, and phase) penetration depths and by explaining when these are relevant. 
The first part of the paper presents a theoretical description that aims to provide physical insight in the origin of the various penetration depths.
It also links them to the optical properties of a microcavity.
The second part presents measurements of the penetration depth in two types of microcavities. 
Measurements on the frequency tuning of the modes in a planar cavity yield the frequency penetration depth.
Measurements on the transverse mode splitting in a plano-concave cavity yields the modal penetration depth.
We compare these two results with each other and with theoretical predictions. 

\section{Optical penetration in DBRs}
\label{sec:DBR}

We consider the reflection of light from a thick, lossless, planar DBR. 
The alternating layers have refractive indices $n_L$ and $n_H$ for the low and high index material and layer thicknesses $d_L$ and $d_H$ such that $n_L d_L = n_H d_H = \lambda_c/4$ for vacuum resonance wavelength $\lambda_c$.
Light is incident from a medium with index $n_{in}$ (typically air with $n_{in} = 1$). 

The most prominent feature of DBRs is the existence of a stopband, or bandgap, which is a frequency range where light cannot propagate and where a thick lossless DBR reflects all incident light. 
The full spectral width of the stopband is \cite{Haus,Yeh}
\begin{equation}
\label{omega-gap}
    \Delta \omega_{gap} = \omega_c \frac{4}{\pi} \arcsin{(\frac{n_H-n_L}{n_H+n_L})} \approx \omega_c \frac{2}{\pi} \frac{\Delta n}{\overline{n}}
\end{equation}
where $\omega_c=2\pi c/\lambda_c$ is the resonance frequency. 
The approximation is valid for small to modest index contrast, where $\Delta n \equiv n_H-n_L \ll \overline{n}$ with average index $\overline{n}=(n_L+n_H)/2$. \cite{Yeh}

At resonance, in the center of the stopband, the forward-propagating field decays exponentially into the DBR, such that its amplitude decreases by a factor $n_L/n_H$ per DBR pair \cite{Yeh, Haus}.
The associated $1/e$ penetration depth $L_I$ of the optical intensity is 
\begin{equation}
    L_I = \frac{1}{2} (\frac{\lambda_c}{4n_L} + \frac{\lambda_c}{4n_H}) \,\ln{(\frac{n_H}{n_L})} \approx \frac{\lambda_c}{4 \Delta n} \,.
\end{equation}
The approximation again applies to the limit of small index contrast, $\Delta n \ll \overline{n}$. 

It seems natural to call $L_I$ ``the penetration depth'' of the DBR and to model the DBR as an effective hard mirror positioned at a distance $L_I$ behind the front surface of the DBR. 
But this is wrong for several reasons.
First and most important, there is no single position at which a hard mirror can mimic all reflection properties of the DBR simultaneously.
Below we will argue that one actually needs three different penetration depths to mimic either (i) the reflection phase, or (ii) the time delay upon reflection, or (iii) the imaging of a focused beam upon reflection. 
Second, these penetration depths depend on the refractive index $n_{in}$ of the incident medium.
Finally, they also depend on whether the DBR starts with a high-index layer (H-DBR) or a low-index layer (L-DBR).
Only if one considers the time delay upon reflection from a H-DBR does one obtain the easy ``natural'' result $L_\tau = L_I$ (see below).

Figure \ref{fig:simulation phase vs freq} shows the calculated frequency dependence of the reflectivity $|r|^2$ and reflection phase $\varphi$ at normal incidence for a typical DBR, similar to the ones used in our experiments.
This figure shows that the reflectivity is approximately constant inside the stopband. 
The interesting physics is contained in the reflection phase $\varphi(\omega)$, which is defined relative to the front facet and scales as $\varphi \propto (\omega-\omega_c)$.
The insets show the physical origin of this phase change:
the node of the standing wave, which resides at the DBR surface at resonance, shifts into ($\varphi > 0$) or out of ($\varphi < 0$) the DBR at frequencies $\omega > \omega_c$ and $\omega < \omega_c$, respectively.
Note the deviations from this linear behavior towards the edges of the stopband, where the maximum shift is approximately half a layer thickness for a H-BDR (see Supplement 1).

\begin{figure}
    \centering
    \includegraphics[width=1\linewidth]{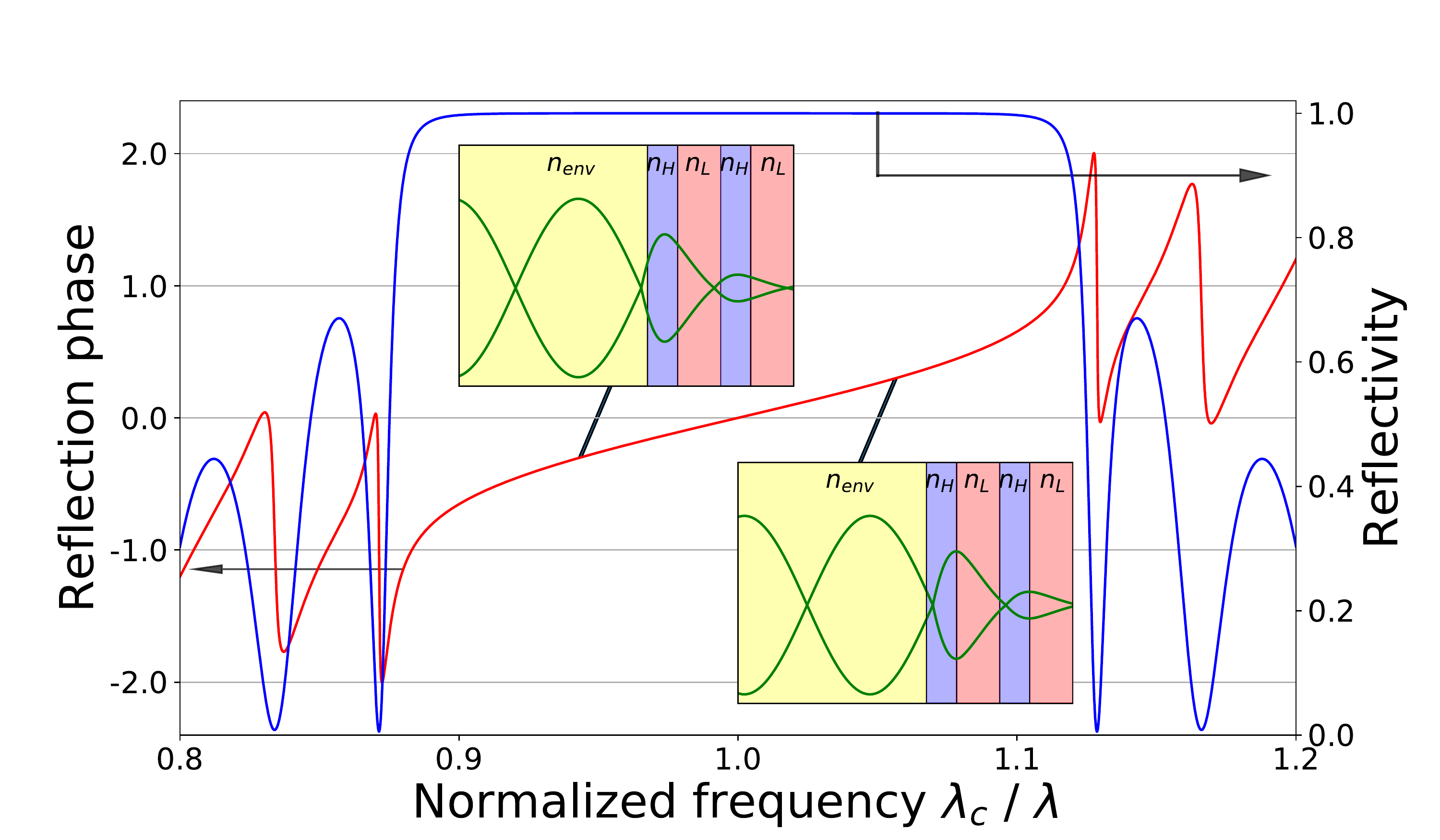}
    \caption{Calculated reflectivity (blue) and reflection phase (red) of DBR versus frequency, normalized to the center frequency. The DBR starts and ends with high-index material, comprises 31 layers with $n_L=1.46$ and $n_H=2.09$, is deposited on a $n_L$ substrate and embedded in air ($n_{in}=1$). The insets sketch how the nodes in the electric field shift into/out of the DBR when the frequency is higher/lower than the center frequency.}
    \label{fig:simulation phase vs freq}
\end{figure}

To describe the reflection of a general (non-monochromatic non-planar) beam of light, we decompose the incident light in its Fourier components.
We label these components by their frequency $\omega$ and transverse wavevector $k_\rho = k_{in} \sin{\theta_{in}}$, where $k_{in} = n_{in} k_0$ is the wavevector in the incident medium and $k_0 = 2\pi/\lambda_0$ and $\lambda_0$ are the wavevector and wavelength in vacuum.
Each monochromatic plane-wave component will reflect with its own reflection amplitude $r(\omega,k_\rho) \exp{[i\varphi(\omega,k_\rho)]}$. 

Inside the stopband, $r(\omega,k_\rho)$ is approximately constant and equal to $r_c \approx \pm 1$. 
The optical field has an anti-node ($r_c = 1$) at the front facet for a L-DBR and a node ($r_c = -1$) for a H-DBR. 
For frequencies near $\omega_c$ and small incident angles, the reflection phase $\varphi(\omega,k_\rho)$ can be approximated by \cite{Babic1992,Babic1993,Ram1995}
\begin{equation}
\label{eq:reflection-Taylor}
\begin{aligned}
    \varphi(\omega,k_\rho) = 2k_{in} L_\varphi & \approx \frac{\partial \varphi}{\partial \omega}  (\omega-\omega_c) + \frac{1}{2} \frac{\partial^2 \varphi}{\partial k_\rho^2}  k_\rho^2 \\ 
    & = 2(k_{in}-k_c) L_\tau -  \frac{k_\rho^2}{k_{in}} L_D \,.
\end{aligned}
\end{equation}
The approximation is a Taylor expansion, where $\partial \varphi/\partial k_\rho = 0$ due to mirror symmetry. 
The final equation defines the frequency penetration depth $L_\tau$ and the modal penetration depth $L_D$ in terms of derivations of the reflection phase. 
We prefer to call $L_\tau$ the frequency penetration depth, whereas others have called it the phase penetration depth \cite{Babic1992}, because our name links $L_\tau$ with frequency tuning (see below).

The frequency penetration depth $L_\tau = c\tau/(2 n_{in})$ quantifies the group delay $\tau = \partial \varphi / \partial \omega$ that an optical pulse experiences upon reflection from a DBR when its optical spectrum fits well within the stopband. 
A hard mirror positioned at a distance $L_\tau$ in the incident medium will produce the same group delay and will hence mimic the time/frequency properties of the DBR.

The modal penetration depth $L_D$ quantifies the imaging properties of the DBR reflection.
A hard mirror positioned at a distance $L_D$ in the incident medium will reflect light with the same angle dependence and will hence produce the same imaging as the DBR. 
Note that the reflection of the DBR depends on angle because the center of the stopband shifts to higher frequencies at non-zero angles of incidence as $\omega_c(k_\rho)-\omega_c(k_\rho=0) \propto k_\rho^2$.
This dependence results in the relation $L_D/L_\tau = \beta = n_{in}^2(n_L^{-2}+n_H^{-2})/2 \approx (n_{in}/\overline{n})^2$ \cite{Babic1992}.
This relation is intuitive, because $L_\tau$ is associated with a time delay, which scales with $\overline{n}/n_{in}$, and $L_D$ is associated with an imaging shift, which scales with $n_{in}/\overline{n}$ (see Supplement 1). With the factor $\beta$ we can rewrite the phase penetration in Eq. (\ref{eq:reflection-Taylor}) depth in terms of $L_\tau$,
\begin{equation}\label{eq:lphi in therm of ltau}
    L_\varphi=\left(\frac{k_{in}-k_c}{k_{in}}-\frac{1}{2}\beta \theta_{in}^2\right) L_\tau.
\end{equation}
This equation shows that small angles $\theta_{in}$ only have a small impact on $L_\varphi$, because $\beta$ is typically small for $n_{in}=1$. 

The phase penetration depth $L_\varphi = \varphi/(2k_{in})$ that we define in Eq. (\ref{eq:reflection-Taylor}) is new in literature. 
We explicitly define this quantity because $L_\varphi$ determines the resonance condition of DBR-based microcavities, rather than $L_\tau$ or $L_D$; see Eqs. (\ref{eq:resonance planar cavity}) and (\ref{eq:concave-resonance}) below. 
$L_\varphi$ also determines the locations of the anti-nodes in the microcavities, where light-matter coupling is maximal.
Equation (\ref{eq:lphi in therm of ltau}) shows that $L_\varphi(\omega)/L_\tau = (\omega-\omega_c)/\omega$ at normal incidence for $n_{in}=1$. 

Babic et al. \cite{Babic1993} have calculated the frequency penetration depth $L_\tau = c\tau/(2 n_{in})$, using transfer matrices. 
Although Babic et al. only analyzed so-called ``matched DBRs'', where all reflections interfere constructively, their results 
\begin{equation}
\label{eq:DBR-slope2}
    \tau = \left(\frac{n_{in}}{n_H}\right) \left(\frac{n_H}{n_H-n_L}\right)  \left(\frac{\pi}{\omega_c}\right) \,\, \mathrm{\text{(H-DBR)}} \quad \mathrm{or} \quad \tau = \left(\frac{n_L}{n_{in}}\right) \left(\frac{n_H}{n_H-n_L}\right) \left(\frac{\pi}{\omega_c}\right)  \,\, \mathrm{\text{(L-DBR)}} 
\end{equation}
also apply to the general case. 
The delay time $\tau$ is different for high-index and low-index DBRs due to the interference of the reflection from the first interface of the DBR with the reflections from the bulk (see Supplement 1). 
We have checked both equations (\ref{eq:DBR-slope2}) with numerical calculations based on transfer matrices (see Supplement 1).
Brovelli et al. \cite{Brovelli1995} have performed similar calculations, using coupled-mode theory for DBRs with small index contrast.
Their results agree with the ones obtained by Babic \cite{Babic1993} in the limit of small index contrast (see Supplement 1).

We like to finish this section by noting that the theory described above is based on several assumption.
First of all, the truncated Taylor expansion in Eq. (\ref{eq:reflection-Taylor}) is valid only for small frequency detunings and small angles.
Second, we have neglected polarization effects. 
These will play a role at larger angles where the Fresnel reflection coefficients depend on polarization. As a result, the spectral width of the stopband will increase for $s$-polarized light and decrease for $p$-polarized light.
Finally, we have neglected dispersion effects. 
For our DBRs and wavelength, the ratio between the group and phase refractive index is $\approx 1.013$ for the SiO$_2$ layers and $\approx 1.058$ for the Ta$_2$O$_5$ layers \cite{rii}. 
The combined effect of dispersion, as a weighted average of these values, results in a modest 3 \% reduction of the spectral width of the stopband and an associated 3 \% increase of the penetration depth. 
We have neglected this effect in our analysis. 

\subsection*{Consequences for cavity resonances}

The resonances of any optical cavity are determined by the condition that the round-trip phase delay is a multiple of $2\pi$. 
For a cavity with two planar DBRs illuminated at normal incidence this results in 
\begin{equation}
    2k_{in} L_{cav} + \varphi_1 + \varphi_2 = n_{in} k_0 (2L_{cav} + 2L_{\varphi 1} + 2L_{\varphi 2}) = q\, 2\pi
    \label{eq:resonance planar cavity}
\end{equation}
where $L_{cav}$ is the distance between the front facets of the two DBRs and $\varphi_1$ and $\varphi_2$ are the reflection phases of the two DBRs (note that $\varphi = 2k_{in} L_\varphi$). 
The longitudinal mode number $q$ counts the number of half wavelengths in the standing wave pattern between the mirrors.
This number $q$ is integer when both mirrors are either H-DBR or L-DBR and hence both have an anti-node or node close to their front facet.
For cavities with one H-DBR and one L-DBR we need to replace $q$ by $q+\frac{1}{2}$ to keep $q$ integer and still account for the sign difference in the reflection $r_c = \pm 1$ of the two mirrors.

Now suppose we change the cavity length and measure the resulting change in resonance frequency/wavelength at fixed $q$. 
We calculate this change by substituting $L_{cav}(k_0)$ in Eq. (\ref{eq:resonance planar cavity}) and taking the derivative of this equation with respect to $k_0$ to arrive at equation (\ref{eq:frequency spacing}).
In the process we use $d \varphi/dk_0 = 2L_\tau /n_{in}$ in the left equation (\ref{eq:resonance planar cavity}) or $L_\varphi(\omega)/L_\tau \approx (\omega-\omega_c)/\omega$ in the middle equation (\ref{eq:resonance planar cavity}).
When we rewrite the end result in terms of $\lambda$ we find a normalized slope
\begin{equation}
    \lambda dL_{cav}(\lambda)/d\lambda = L_{cav} + L_{\tau 1} + L_{\tau 2}  \,,
    \label{eq:frequency spacing}
\end{equation}
This equation shows that the relevant penetration depth for a frequency scan is $L_\tau$ and not $L_\varphi$. 
The combination $L_{cav} + L_{\tau 1} + L_{\tau 2}$ also determines the quality factor $Q$ of the optical resonances and the associated cavity loss rate $\omega_0/Q$.

The resonance condition of plano-concave cavities differs from that of planar cavities by the so-called Gouy phase. 
For planar-concave cavities with hard mirrors, the resonant modes are Hermite-Gaussian TEM$_{n,m}$ modes with flat wavefronts at the planar mirror and matched curved wavefronts at the concave mirror. 
Upon propagation, these modes experience a phase lag relatively to a plane wave.
This phase lag is proportional to $(n+m+1)$ and to the Gouy phase $\theta_{Gouy}=\arcsin(\sqrt{L/R})$ of the fundamental TEM$_{00}$ mode, where $L$ and $R$ are the cavity length and mirror radius.

The resonant optical modes of a planar-concave cavity with DBR mirrors are also TEM$_{n,m}$ modes.
Their (round-trip) resonance condition 
\begin{equation}
\label{eq:concave-resonance}
    n_{in} k_0 (L_{cav}+L_{\varphi 1}+L_{\varphi 2}) - (n+m+1) \theta_{Gouy} = q\, \pi \,,
\end{equation}
again includes the phase penetration depth $L_\varphi$ of both mirrors. 
It also includes a Gouy phase that is now given by $\theta_{Gouy}=\arcsin(\sqrt{(L_{cav}+2L_D)/R})$.
Note that the relevant penetration depth in this equation is $L_D$, because the Gouy phase is associated with phase fronts and thereby linked to imaging properties. 
We assume that the value of $L_D$ for the curved mirror and the flat mirror are the same, and therefore use $2 L_D$.
The radius of curvature $R$ is an effective radius, which includes all small variations of the mirror curvature in the fabrication process \cite{Trichet2015}. 
The Gouy phase co-determines the Rayleigh range and waist of the cavity modes and hence also the optical mode volume and attainable atom-field interaction. 

\subsection*{Comparison with literature}

A comparison of our results with literature shows that the subtleties of multiple penetration depths are often overlooked. 
We will give a few examples of how equations would have been different if our theory would have been used.  

First, our Eqs. (\ref{eq:resonance planar cavity}) and (\ref{eq:concave-resonance}) show that the resonance condition depends on the phase penetration depth $L_\varphi$, which is typically small and zero at resonance. 
But Eqs. (2) and (3) in ref. \cite{Trichet2014} state that the resonance condition contains a single wavelength independent penetration depth. The authors later use the same quantity to describe the frequency tuning of the resonances, whereas our Eq. (\ref{eq:frequency spacing}) shows that frequency tuning depends on the frequency penetration depth $L_\tau$.

Second our Eqs. (\ref{eq:resonance planar cavity}) and (\ref{eq:frequency spacing}) show that the frequency spacing between consecutive longitudinal modes depends on the frequency penetration depth $L_\tau$.
But the authors in \cite{Greuter2014} determine the longitudinal mode value by taking $q=2 \frac{\partial L_{cav}}{\partial \lambda}$ and hence forget the contribution of $L_\tau$. 

Third, our Eq. (\ref{eq:concave-resonance}) shows that the frequency spacing between the transverse modes depends on the modal penetration depth $L_D$. 
Equation (3) in ref. \cite{Trichet2014} uses the expression $\theta_{Gouy} = \arccos{\sqrt{1-L_{cav}/R}}$ for the Gouy phase and hence does not take any penetration into account. 
Reference \cite{Greuter2014} makes the same mistake in their Eq. (1). 
Reference \cite{Benedikter2019} uses the frequency penetration depth in their Eq. (2) for the Gouy phase. 
But the correct equation should have been $\theta_{Gouy}=\arccos{\sqrt{1-(q\lambda/2+2L_D)/R}}$ at $L_\varphi = 0$.

Finally, incorrect use of the penetration depths also affects the Purcell effect.
Our analysis shows that the Purcell factor $F_P$ depends primarily on the modal penetration depth $L_D$, as the increase in mode volume due to the field penetration into the DBRs is compensated by a similar increase of the quality factor.
The effect of $L_D$ on the Purcell factor is typically small but can still become important when the cavity length is order $\lambda$. 
For the $q=2$ fundamental mode in a cavity with one L-DBR, we predict that the modal penetration depth leads to an increase in mode area and an associated decrease of the Purcell factor $F_P \sim w_0^2$ by a factor $1/\sqrt{2}$.  
The consequence of using $L_\tau$ instead of $L_D$ is an underestimation of the Purcell enhancement \cite{Vogl2019}. 

\section{Methods}

\begin{figure}
    \centering
    \includegraphics[width=0.8\linewidth]{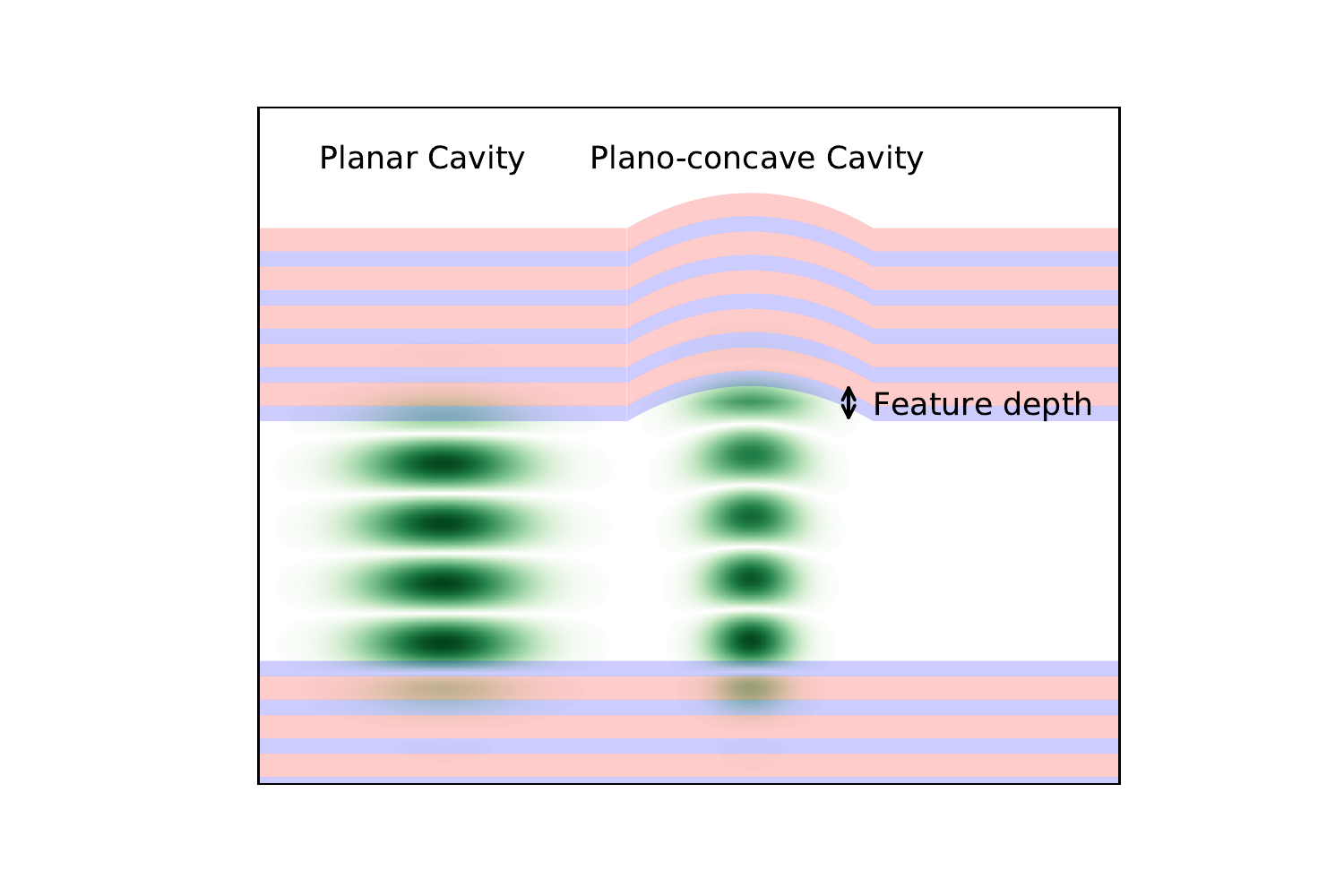}
    \caption{Illustration of the planar and plano-concave cavity modes. The blue and red areas correspond to the high and low index materials of the DBR, with a flat mirror at the bottom and a patterned mirror at the top. The green areas indicate standing waves in the cavity. To excite the planar or plano-concave cavity we focus light either on the flat or on the curved part of the patterned mirror. The indentation at the patterned mirror is referred to as "feature depth". }
    \label{fig:experimental setup}
\end{figure}

%% Removed, because similar information is described in the final paragraph of this subsection
%% The precise determination of the penetration depth is shown below. We present two experiments. First, we measure the wavelength-dependent transmission of the planar cavity for various cavity lengths. Then we measure at which relative mirror positions the plano-concave cavity transmits light with a (monochromatic) laser.  Figure \ref{fig:experimental setup} shows the flat and patterned mirror that we use for the experiments.   

Our planar and patterned mirrors were produced by Oxford High-Q. 
The patterned substrate is fabricated with a focused-ion-beam-etching technique that creates a series of high-quality concave structures with typical radii of curvature of 2-20 $\mu$m  \cite{Trichet2014,Trichet2015,Trichet2018}.
The SiO$_2$ substrates were coated with 31 and 35 alternating layers of SiO$_2$ and Ta$_2$O$_5$, to produce two DBRs that both end with high-index material and hence have virtually identical reflection properties. 
These DBRs have a stopband with a width $\Delta \lambda\approx150$ nm centered around $\lambda_c=640$ nm, as expected for a DBR with $n_L \approx 1.46$ and $n_H \approx 2.09$. 
The transmission of the patterned mirror is $(3.4\pm 0.2) \times 10^{-5}$ and the transmission of the flat mirror is $(1.1 \pm 0.1) \times 10^{-4}$ inside the stopband. 
The transmission through the plano-concave cavity is only $\approx$ 1\%, due to scattering losses on the mirrors. 
These losses are not relevant for the analysis presented in this paper.

One of the mirrors is fixed while the other mirror can be moved with 6 degrees of freedom on a hexapod system.
We align the mirrors to the point where they are parallel and touch each other. This point is referred to as 'touch down'. 
We scan the mirror position from touch down to over >2 $\mu$m distance with sub-nm precision.

The non-linearity of the piezo scan and the point of touch down are determined by measuring the microcavity transmission outside the stopband with a green laser ($\lambda = 520$ nm). 
A CCD image of the microcavity confirms the parallelism when no fringes are visible. 
The point of touch down is determined from the part in the scan where the transmitted intensity is constant.

In the next section, we will present accurate measurements of penetration depths on a planar and a plano-concave microcavity, as indicated in the section titles. 
Figure \ref{fig:experimental setup} shows that we use the same mirrors in both experiment, but we focus light on different parts. 
The planar microcavity transmission spectra presented in Fig. \ref{fig:phase penetration depth} are obtained with a spatially-filtered Xenon lamp and a fiber-coupled spectrometer. 
The experimental results on the transverse mode splitting of the plano-concave microcavity, presented in Figs. \ref{fig:transverse mode splitting cav2a} and \ref{fig:results:Lpenvsrad}, are obtained by measuring the microcavity transmission of a HeNe laser ($\lambda = 633$~nm) with a photo-multiplier tube.
The light was coupled in with an $f=7.5$ mm lens and coupled out with an $f=8$ mm lens.

\section{Results}

\subsection{Frequency penetration depth $L_\tau$ (planar cavity)}

\begin{figure}
\label{fig:fig3}
\centering
    \includegraphics[width=1\linewidth]{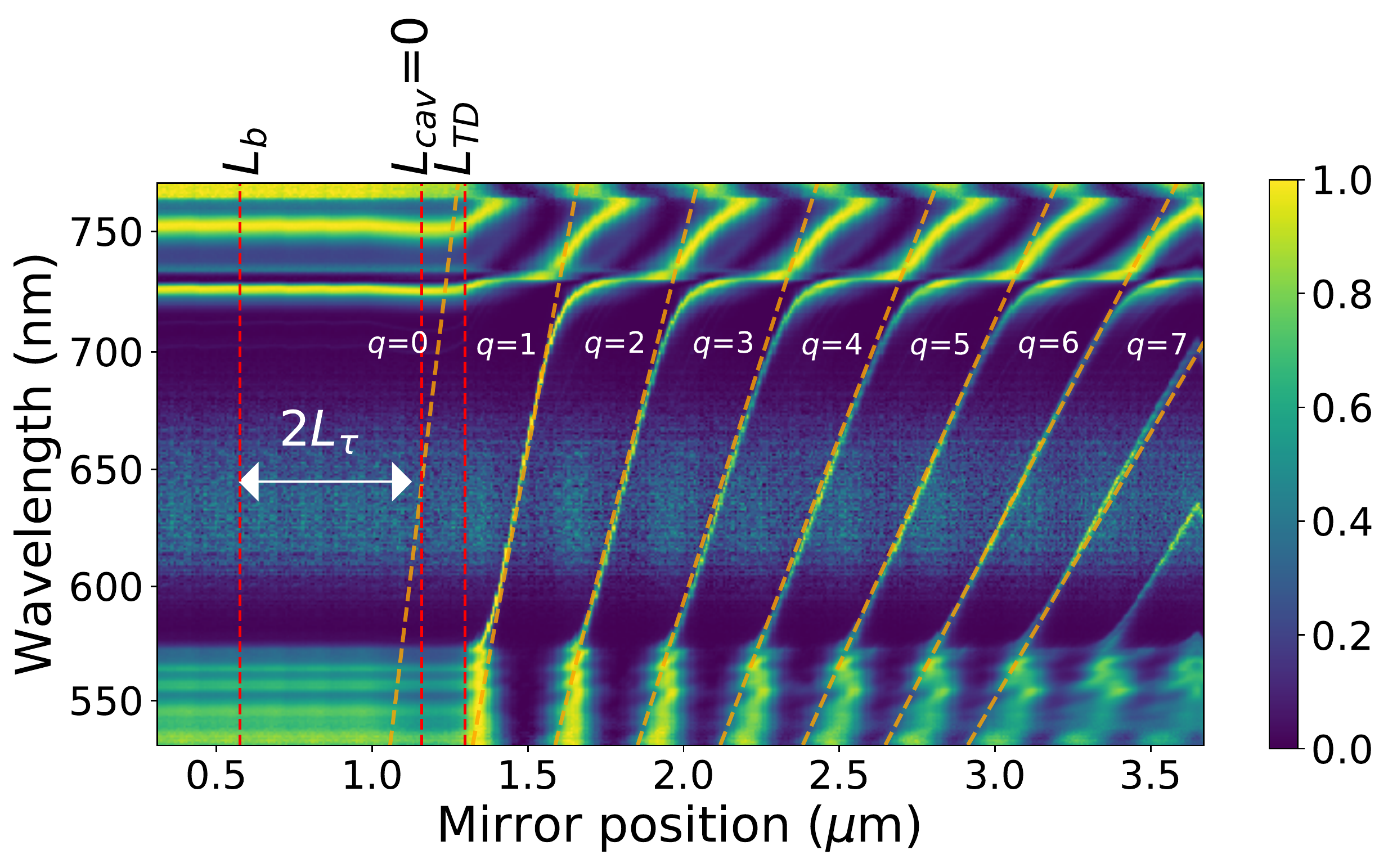}
    \caption{False-color plot of the normalized transmission spectrum for a planar cavity. We scan the mirror position from 1.0 to >3.6 $\mu$m, which includes the point of 'touch down' (vertical line labeled $L_{TD}$) and repeat the spectra below $1.0~\mu$m for aesthetic reasons. The slanted lines in the central region of the spectrum show the planar cavity modes in the stopband. These modes are labeled by their mode number $q$ and are fitted with straight (dashed orange) lines. We add the calculated $q=0$ mode, which per definition intersects the vertical $L_{cav}=0$ at the central wavelength $\lambda_c$. All fits are constrained to cross $\lambda=0$ at the same mirror position $L_b$ (see text).}
    \label{fig:phase penetration depth}
\end{figure}

In the first experiment, we measure the transmission spectrum $P(\lambda;L)$ of the planar cavity.
For each wavelength $\lambda$, the transmitted power varies between $P_{min}(\lambda)$ and $P_{max}(\lambda)$ with $L$. 
We use these extrema to normalize the transmission spectrum between $0$ and $1$ and show the results as false-color plot in Fig. \ref{fig:phase penetration depth}. 
Due to this normalization, this figure does not show the 4 orders of magnitude difference between the very low transmission ($10^{-4}$) for wavelengths inside the stopband and the order unity transmission outside the stopband. 

The cavity length is varied from just below touch down (red dashed line indicated by $L_{TD}$) to a mirror position $\approx$ 3.6 $\mu$m. Below the point of touch down, the cavity length is constant and close to zero. This part is only included to show the derived quantities $L_{cav}=0$ and $L_{b}$ (see below).  

The slanted lines in the central region of the spectrum show the planar-cavity modes in the stopband.
These lines become non-linear towards the edges of the stopband, where the reflection phase $\varphi$ makes a phase jump, in agreement with the theoretical Fig. \ref{fig:simulation phase vs freq}.
The planar-cavity modes are labeled by their longitudinal mode number $q$.

The slanted dashed lines in Fig. \ref{fig:phase penetration depth} result from a simultaneous constrained linear fit of the $q=1-7$ longitudinal cavity modes in the linear part of the stopband (600-680 nm). 
This fit is heavily constrained and contains only one fit parameter, $L_b$, because all fit lines are $\lambda/2$ apart and hence cross at the same point $L_b$ for $\lambda = 0$.
Small deviations from the fit lines observable for $q=2-5$ originate from an imperfect correction of the non-linearity of the piezo scan.
These fits also allow us to extrapolate to the virtual $q=0$ mode.
At the center wavelength, where $\varphi=0$, the $q=0$ line coincides with the point $L_{cav}=0$, indicated by the middle red dashed line.

The key result in Fig. \ref{fig:phase penetration depth} is the observation that the $q=0$ mode is also slanted or, equivalently, that the point $L_b$ does not correspond to $L_{cav}=0$. 
The distance between these points, indicated by the arrow in Fig. \ref{fig:phase penetration depth}, yields the frequency penetration depth $L_\tau = 0.28 \pm 0.02~\mu$m; see Eqs. (\ref{eq:resonance planar cavity}) and (\ref{eq:frequency spacing}) for theory.
The uncertainty estimate is based on a comparison between results from different measurement series and different methods of analysis, both manually and by computer.  
Note that the analysis presented above was based on the reasonable assumption that the first mode after touch down is the $q=1$ mode. 
This assumption yields a distance $0.14 \pm 0.02~\mu$m between touch down and zero cavity length.
If the first mode would have been $q=2$, this would have led to a much larger distance of 0.46$\pm$0.02 $\mu$m and an unrealistically low value of $L_\tau=0.12\pm0.02~\mu$m (see discussion below).

\subsection{Modal Penetration depth $L_D$ (plano-concave cavity)}

In the second experiment, we measure the transmission of a HeNe laser through a plano-convave cavity while scanning the cavity length. 
Each group of transmission peaks contains the fundamental TEM$_{00}$ mode and multiple high-order TEM$_{nm}$ modes. 
The wavelength of the HeNe ($\lambda=633$ nm) is close enough to the center wavelength ($\lambda_c=640$ nm) to neglect the phase penetration depth (theory predicts $2L_\varphi\approx -0.01 ~\mu$m).

\begin{figure}
\label{fig:fig1}
    \centering
    \includegraphics[width=1\linewidth]{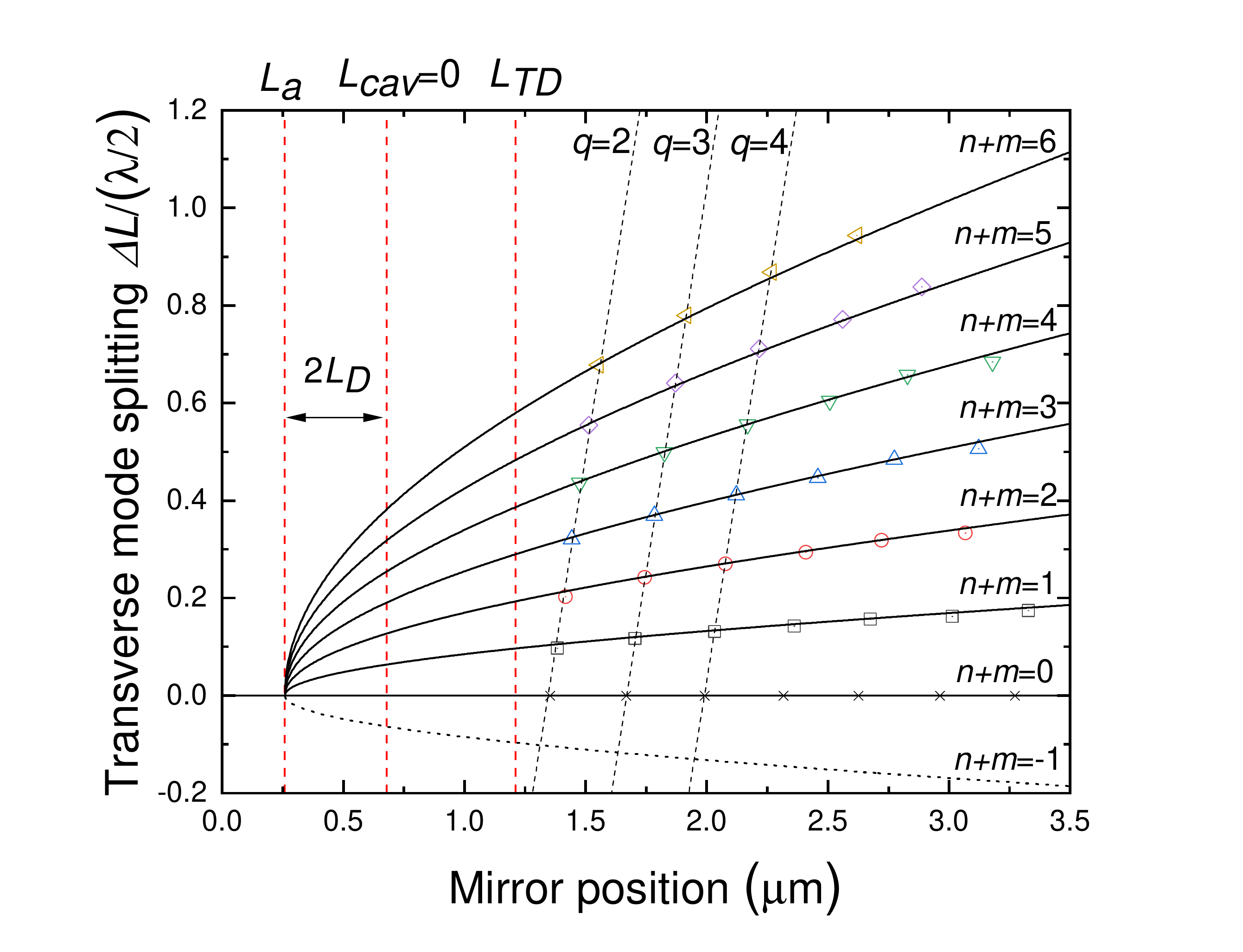}
    \caption{Transverse mode splitting versus mirror position $L$ in a plano-concave cavity. This mode splitting is expressed as the displacement $\Delta L$ between the resonance of the high-order mode ($n+m=1-6$) and the corresponding fundamental mode ($n+m=0$). The combined fit of these data, depicted as a set of black curves, yield the radius of curvature and a fictitious mirror position $L_a$ where all transverse modes are frequency degenerate. We add the fictitious cavity mode ($n+m=-1$) to compare with the planar cavity and to find $L_{cav}=0$ (see text).}
    \label{fig:transverse mode splitting cav2a}
\end{figure}

Figure \ref{fig:transverse mode splitting cav2a} shows the measured splitting $\Delta L$ between each transverse higher-order mode (indicated by $n+m>0$) and the associated fundamental mode ($n+m=0$) as a function of mirror position. 
We measured these splittings for 7 groups of modes, of which the first three ($q=2,3,4$) are indicated by dashed black lines. Below, we will explain why we start counting from $q=2$. 

The solid curves are based on a simultaneous fit of all measurements using two fit parameters: the mirror radius $R$ and the position $L_a$ of full degeneracy of the transverse modes. 
Our fit yields $R = 10.7\pm0.1 ~\mu$m and $L_a = 0.26 \pm 0.03~\mu$m (indicated by the left red dashed line). 
Figure \ref{fig:transverse mode splitting cav2a} shows that these estimates require a serious extrapolation of the data.
The computer-generated error bars are based on statistical errors only and might thus be optimistic (0.03 $\mu$m statistical error in $L_a$) as they do not take systematic errors into account.
A possible systematic error could be a deviation of the transverse mode splitting, and the associated Gouy phase, from the simple paraxial theory \cite{Luk1986}. 
In the absence of an alternative theory, we cannot estimate the size of this systematic errors. 
From an experimental point of view, we can only determine statistical errors to find that errors of multiple measurements on multiple cavities agree with each other (see below).

In Fig. \ref{fig:transverse mode splitting cav2a}, we have added an extra (black dotted) curve for the virtual $n+m=-1$ modes.
A comparison of Eqs. (\ref{eq:resonance planar cavity}) and (\ref{eq:concave-resonance}) shows that these virtual modes should have the same resonances as the planar cavity modes. 
By extrapolation of these virtual planar modes to $q=0$ we find the point $L_{cav}=0$, indicated as the middle dashed line.

The key result in Fig. \ref{fig:transverse mode splitting cav2a} is the modal penetration depth. 
Equation (\ref{eq:concave-resonance}) shows how this value can be obtained from the distance $2L_D$ between the leftmost vertical lines, assuming identical penetration depth in the flat and curved mirror.
From the analysis of Fig. \ref{fig:transverse mode splitting cav2a} we thus obtain a measured modal penetration depth $L_D = 0.211 \pm 0.015 ~\mu$m. 

Finally, we note that the distance between touch down and $L_{cav}=0$ in Fig. \ref{fig:transverse mode splitting cav2a} is $0.53 \pm 0.03~\mu$m.
This value is larger than for the planar cavity because it contains the feature depth of the concave mirror (see Fig. \ref{fig:experimental setup}). 
Trichet et al. \cite{Trichet2015} report feature depths of $0.30 ~\mu$m and $0.32 ~\mu$m for curved mirror radii of $23 ~\mu$m and $12 ~\mu$m, respectively. 
By comparing transmission spectra of planar and plano-concave cavities, we find $0.34\pm0.02 ~\mu$m and $0.35\pm0.02 ~\mu$m for our mirrors.
With a feature depth of $\approx \lambda/2$, the lowest $q$ mode of the plano-concave cavity will be $q=2$, while the planar cavity has $q=1$. 
After subtraction of the measured feature depth, we determine the spacing between the planar parts of the mirrors to be $0.18 \pm 0.03~\mu$m at touch down in this new alignment.  

\begin{figure}
\label{fig:fig2}
    \centering
    \includegraphics[width=\linewidth]{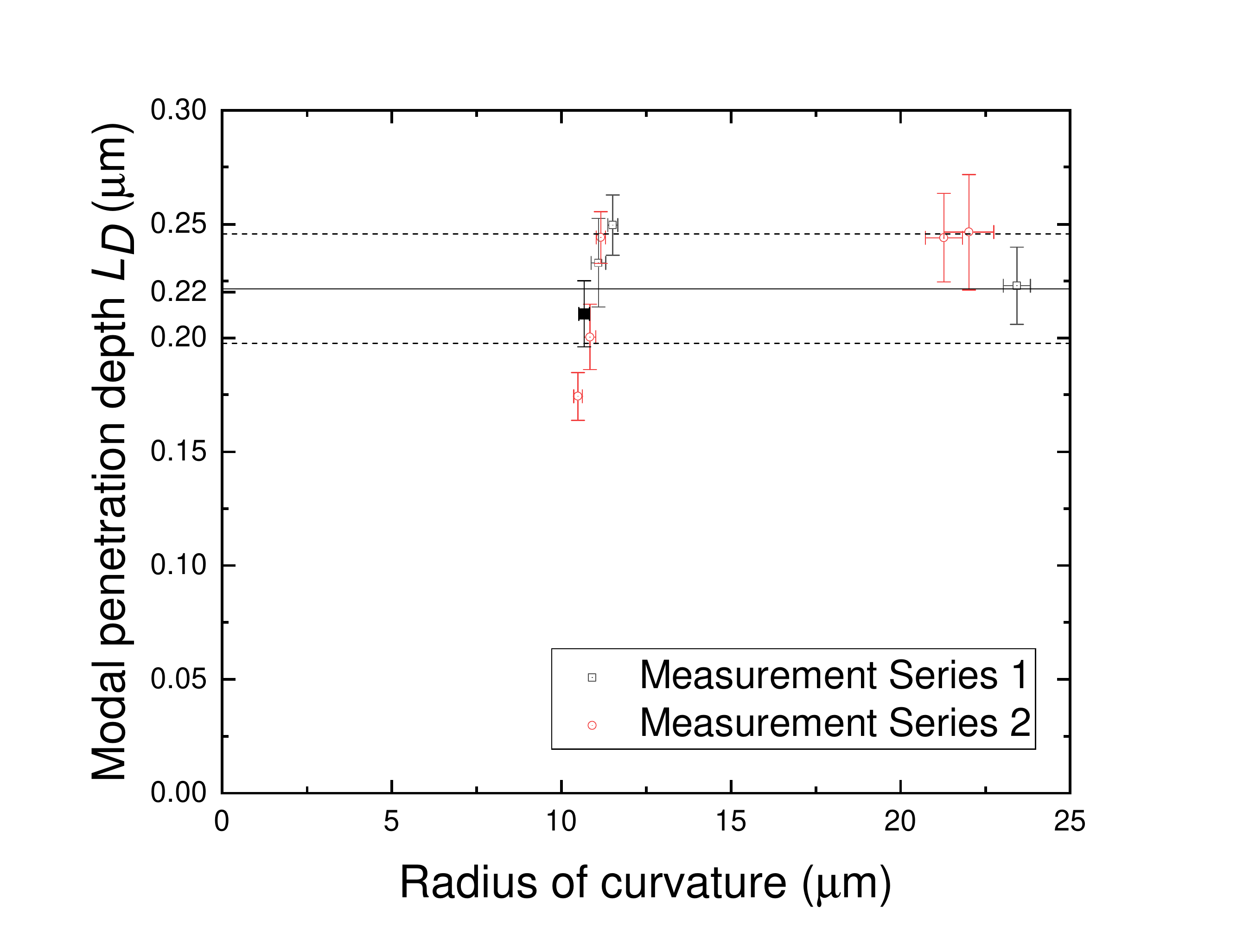}
    \caption{Modal penetration depth and radii of curvature for nine measures on various plano-concave cavities. The solid point results from Fig. \ref{fig:transverse mode splitting cav2a}. The red and black points are obtained in two measurement series. The horizontal lines show the averaged modal penetration depth and the error range, corresponding to $\overline{L_D}=0.22\pm0.02~\mu$m.}
    \label{fig:results:Lpenvsrad}
\end{figure}

We performed the analysis depicted in Fig. \ref{fig:transverse mode splitting cav2a} on 9 data sets, obtained from 6 different cavities on 2 different days (3 cavities were measured on both days). 
We only analyzed data sets that contained at least four clearly visible transverse modes.  
The solid point in Fig. \ref{fig:results:Lpenvsrad} shows the fit parameters obtained from Fig. \ref{fig:transverse mode splitting cav2a}.
The colors of the points indicate measurement series on different days. 

%% The analysis depicted in Fig. \ref{fig:transverse mode splitting cav2a} was done 9 times in total with different cavities on different days. The solid point in Fig. \ref{fig:results:Lpenvsrad} shows the fit parameters obtained from Fig. \ref{fig:transverse mode splitting cav2a}. The colors of the points indicate measurement series on different days, where the cavity incoupling is slightly different. Three cavities were analyzed in both series, three other cavities were analyzed either in series 1 or 2. We only analyzed data sets that contained at least four clearly visible transverse modes. 

The data points in Fig. \ref{fig:results:Lpenvsrad} are divided in two groups, corresponding to cavities with $R = 10-11~\mu$m and $R = 21-23~\mu$m. 
The distribution of the data points shows that the modal penetration depth is approximately the same for all cavities and does not depend on mirror radius over the studied range.

The horizontal lines show the weighted average of the modal penetration depth with its intrinsic error $\overline{L_D}=0.22\pm0.02~\mu$m. 
This intrinsic error is based on the spread in the measurements, which is slightly larger than the error bars estimated for individual measurements. 
This estimate only contains statistical errors. 

\section{Discussion}

We start the discussion by comparing experiment with theory.  
We have measured a frequency penetration depth $L_\tau=0.28 \pm 0.02~\mu$m and modal penetration depth $L_D=0.22 \pm 0.02~\mu$m. 
For our H-DBRs, with $n_L =1.46$ and $n_H=2.09$, we predict $L_\tau=0.25~\mu$m and $L_D= 0.09~\mu$m. 
The measured frequency penetration depth $L_\tau$ is in reasonable agreement with theory.
The measured modal penetration depth $L_D$ is smaller than $L_\tau$, also as expected.
But the measured value of $L_D$ is not as small as theory predicts.

The observed discrepancies in $L_D$ are most likely due to simplifications in the theory. 
We will mention four and start with the least likely.
The most crucial simplification seems to be the Taylor expansion of the reflection phase in Eq. (\ref{eq:reflection-Taylor}), which is based on the assumption that all Fourier components fit well within the stopband.
At large angles of incidence the blue shift of the resonance frequency could bring us into the non-linear regime of the reflection phase, where an additional cubic term increases the effects (see refs. \cite{Hood2001,Thorpe2005}).
This scenario sounds reasonable, as the opening angle of the $q=2$ modes is as large as $\approx$ 0.25 rad ($e^{-2}$) for the fundamental mode and much larger for the high-order modes. 
However, even opening angles as large as 0.6 rad will shift the resonance frequency only $\Delta \omega_c/\omega_c \approx (1/2)\beta\theta_{in}^2 \approx 5$ \% or $\approx$ 30 nm, which does seem to be enough to reach the non-linear regime.
A second simplification lies in the treatment of the curved DBR as a mirror with a single curvature \cite{Trichet2015} that mimics the angle-dependent reflection of a flat DBR and hence has the same optical penetration.
Again, this argument is not convincing as the range of angles of incidence on the curved and flat DBR are not very different, given the large opening angle of the cavity modes. 

A simplification that could be crucial is the scalar treatment of the optical field. 
Computer simulations show that polarization effects will become important at larger angles, even far below the Brewster angle. 
Hence, one might have to distinguish transverse modes based on their optical polarization.
Finally, we didn't take potential coating inhomogeneities and thickness distortions of the mirrors into account \cite{Trichet2015,Benedikter2019}.
Small-scale distortions are likely to average out over the mode profile.
But large-scale distortions can deform the transverse modes away from the ideal spherical case and hence result in an incorrect assignment of the radius $R$ and the modal penetration depth $L_D$. 
Furthermore, non-paraxial propagation effects could also lead to an incorrect assignment and introduce systematic errors in the analysis. 
A discussion of all these complications is beyond the scope of the paper.

In conclusion, we have presented an analysis of optical penetration in DBRs and accurate measurements thereof. 
Our analysis shows that there are actually three penetration depths which are relevant in different experiments.
We have measured the frequency penetration depth $L_\tau$ to find that it agrees with theory.
We have also measured the modal penetration depth $L_D$ to find that it is smaller than $L_\tau$, but not as small as expected.
We attribute the observed deviations to theoretical simplifications.
Maybe most important, we have argued that the effect of optical penetration on microcavity resonances is often misinterpreted.
The absolute resonance conditions depend on the reflection phase $\varphi$ and hence on the phase penetration depth $L_\varphi$. 
The frequency spacing between the longitudinal modes and their quality factor depend on the frequency penetration depth $L_\tau$.
The frequency spacing between the transverse modes and their area/cross sections depend on the modal penetration depth $L_D$.
The Purcell factor $F_P$ depends primarily on the modal penetration depth, as the increase in mode volume due to the field penetration into the DBRs is compensated by an increase of the quality factor.

\section*{Acknowledgments}

The authors thank A. A. P. Trichet from Oxford HighQ for supplying the two DBRs, D.I. Babic for helping us to interpret some equations in his papers, and M.J.A. de Dood for his suggestions to improve the manuscript. 

\section*{Disclosures}
The authors declare no conflicts of interest.
\\
\\
See Supplement 1 for supporting content.

%%%%%%%%%%%%%%%%%%%%%%% References %%%%%%%%%%%%%%%%%%%%%%%%%

%%%%%%%%%% If using BibTeX:
\bibliography{sample}

\end{document}